\documentclass[aps,prl,twocolumn,showpacs,groupedaddress]{revtex4}
\usepackage{amsfonts}
\usepackage{amsmath}
\usepackage{amssymb}
\usepackage{graphicx}

\begin{document}
\title{Interplay between sheet resistance increase and magnetotransport properties in ${\rm LaAlO_3}/{\rm SrTiO_3}$}
\author{Snir Seri}
\author{Moty Schultz}
\author{Lior Klein}
\affiliation{Department of Physics, Nano-magnetism Research Center,
Institute of Nanotechnology and Advanced Materials, Bar-Ilan
University, Ramat-Gan 52900, Israel}


\begin{abstract}
We find that the sheet resistance ($R_s$) of patterned samples of
${\rm LaAlO_3}/{\rm SrTiO_3}$ with a length scale of several microns may increase significantly at low temperatures
in connection with driving electrical currents and applying in-plane
magnetic fields. As the samples are warmed up, $R_s$ recovers to
its original value with accelerated recovery near ${\rm 70 \ K}$
and ${\rm 160 \ K}$. Concomitantly with the increase in $R_s$, the
carrier density and the mobility decrease and magnetotransport
properties which may be linked to magnetism are suppressed.

\end{abstract}

\maketitle

\setcounter{secnumdepth}{2}
\section{Introduction}

The ${\rm LaAlO_3}/{\rm SrTiO_3}$ (LAO/STO) system is a
fascinating manifestation of electronic reconstruction at the
interface of two insulating oxides which leads to the emergence of
electrical conductivity with reduced dimensionality. Consequently,
this system has become the focus of intensive experimental and
theoretical effort to elucidate its rich electronic properties
\cite{high mobility,tunable
quasi,superconductivity1,superconductivity2,origin unusual
final,origin perspectives,Mapping,complementary,LAO STO,tuning
spin orbit,rashba spin orbit,electronics,Capacitance}, which is a
challenging task  considering the sensitivity of this system to
many parameters such as oxygen pressure during growth, the
thickness of the LAO layer, post growth treatment, etc.



One of the most intriguing properties of the LAO/STO system is the
emergence of magnetism, which appears to be sensitive to growth
conditions. In LAO/STO samples grown in high oxygen pressure
($P_{O_{2}}\sim 10^{-3}$ mbar), magnetism has been suggested based
on resistivity hysteresis during magnetic field sweeps
\cite{ferromagnetism,Coexistence1}. For samples grown in higher
oxygen pressure ($P_{O_{2}}\sim 10^{-2}$ mbar), based on
magnetization measurements, ferromagnetism was found to persist up
to room temperature \cite{Ariando}.

In LAO/STO samples grown in lower oxygen pressure ($P_{O_{2}}\sim
10^{-4}$-$10^{-5}$ mbar), magnetism has been first suggested based
on magnetotransport properties
\cite{antiferromagnetism,antisymmetry}. Recently, local imaging
using a scanning superconducting quantum interference device
\cite{Coexistence3} and high-resolution magnetic torque
magnetometry measurements \cite{Coexistence2} have provided direct
evidence for magnetic order in such samples. In the latter,
magnetization was found to persist up to $ 200 \ {\rm K}$. The
emerging picture from the local probe measurements is of submicron
patches of ferromagnetism with a wide distribution of size and
easy axis orientation. Interestingly, fewer ferromagnetic
patches were found in a patterned sample \cite{Coexistence3}.

The results presented here suggest that linked to a resistivity
increase, the magnetism may be suppressed post fabrication.

We study patterned LAO/STO samples grown in an oxygen atmosphere of
${\rm 7\times10^{-5}}$ mbar and show that for patterns with length scales of several microns the
sheet resistance ($R_s$) at low temperatures may increase significantly in connection with driving electrical currents and applying in-plane magnetic fields. The
$R_s$ increase is nonuniform; therefore, it yields variations in
the properties of the LAO/STO system within a given sample and
between samples. The initial as-cooled $R_s$ is recovered by warming
the samples. We identify two temperatures ($ 70 \ {\rm K}$ and
$ 160 \ {\rm K}$) near which an accelerated recovery of $R_s$
is observed. We suggest charge trapping as a possible origin of the
$R_s$ increase.

Concomitantly with the $R_s$ increase, the carrier density and the
mobility decrease and there are dramatic changes in the
magnetotransport properties of the samples. As we will show, Hall
effect (HE) and magnetoresistance (MR) properties which suggest the
existence of magnetism in the as-cooled samples are gradually
suppressed as $R_s$ increases.

The observed $R_s$ increase, which may affect dramatically the transport properties of patterns on the order of several microns and less, has important implications for the future design of submicron devices. In addition, the interplay with magnetotransport properties linked to magnetism may help elucidate the intriguing nature of magnetism in this system.


\section{Experiment}

\begin{figure}[ht]
\includegraphics[scale=0.49, trim=100 0 100 0]{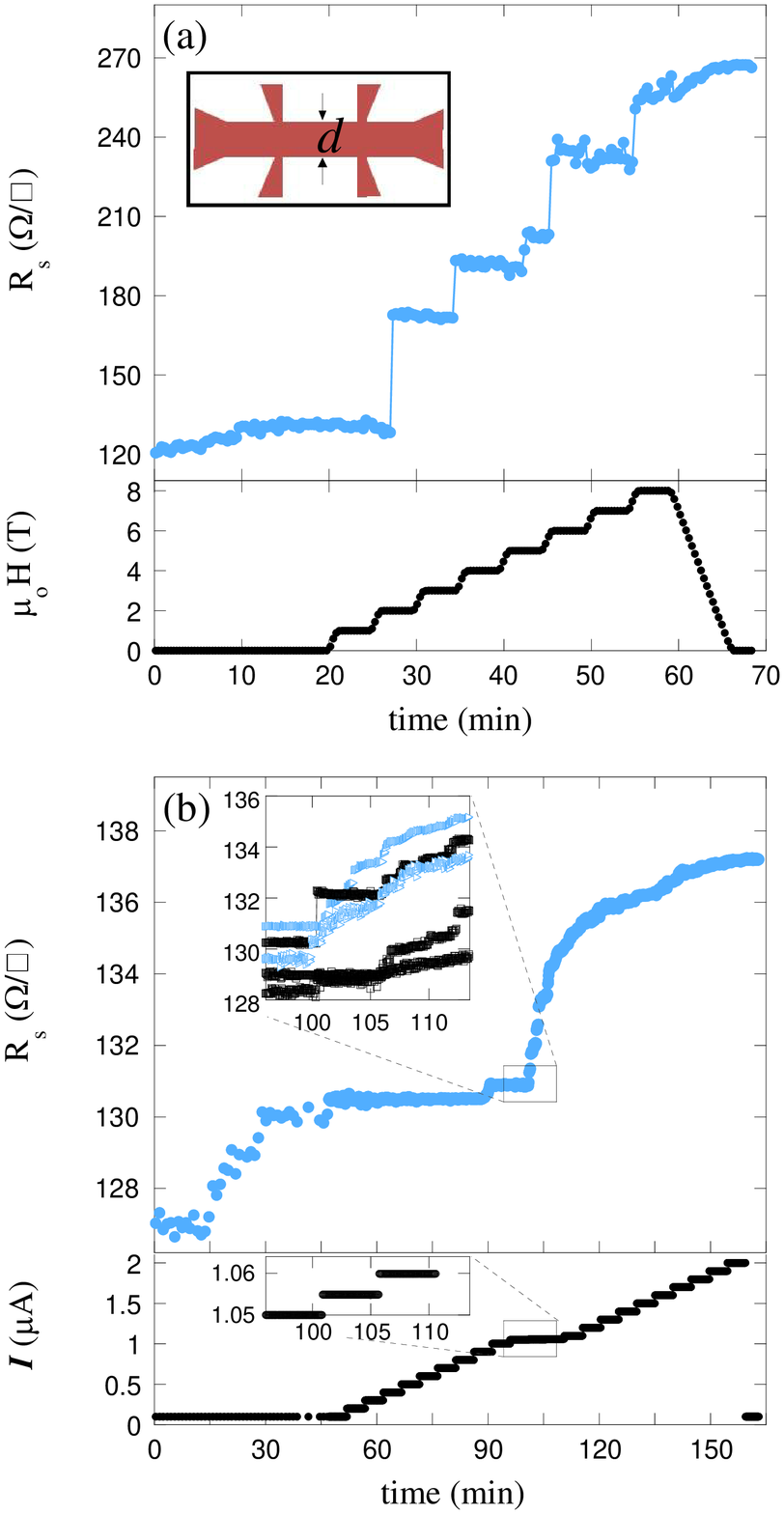}
\caption{(a) $R_s$ as a function of time at 5 K in a pattern with $d=5 \ \mu$m as an in-plane magnetic field is increased in steps. Inset: A sketch of a typical pattern. (b) $R_s$ as a function of time at 5 K in a similar pattern as the current is increased in steps with smaller steps near 1 $\mu$A as shown in the inset for more patterns (black squares) and different cooling cycles of the same pattern (blue triangles).}
\label{trapping}
\end{figure}

The samples were grown by pulsed laser deposition in an oxygen atmosphere of ${\rm 7\times10^{-5}}$ mbar on ${\rm TiO_2}$ terminated (001) STO surfaces at $770^{\circ}$ C. The LAO thickness is four or eight unit cells. The samples were cooled to room temperature in 400 mbars of $\rm O_2$, including a 1-h oxidation step at $600^{\circ}$ C. The laser fluence was about 0.8 J/cm$^2$, with a repetition rate of 1 Hz and a growth rate of about 18-20 shots/unit cell. Patterning was done by photolithography as described in Ref. \cite{microlithography}. The typical geometry of our samples is shown in Figure \ref{trapping}a
(inset). The width of the current path ($d$) in our patterns varies between 2 and 100 $\mu$m. In experiments where the current magnitude is not indicated, we use low currents that do not exceed 0.05 $\mu$A per 1 $\mu$m of wire width. The contact arrangement allows for simultaneous longitudinal and transverse voltage measurements.


\section{Results and Discussion}

\begin{figure}[ht]
\includegraphics[scale=0.49, trim=100 0 100 0]{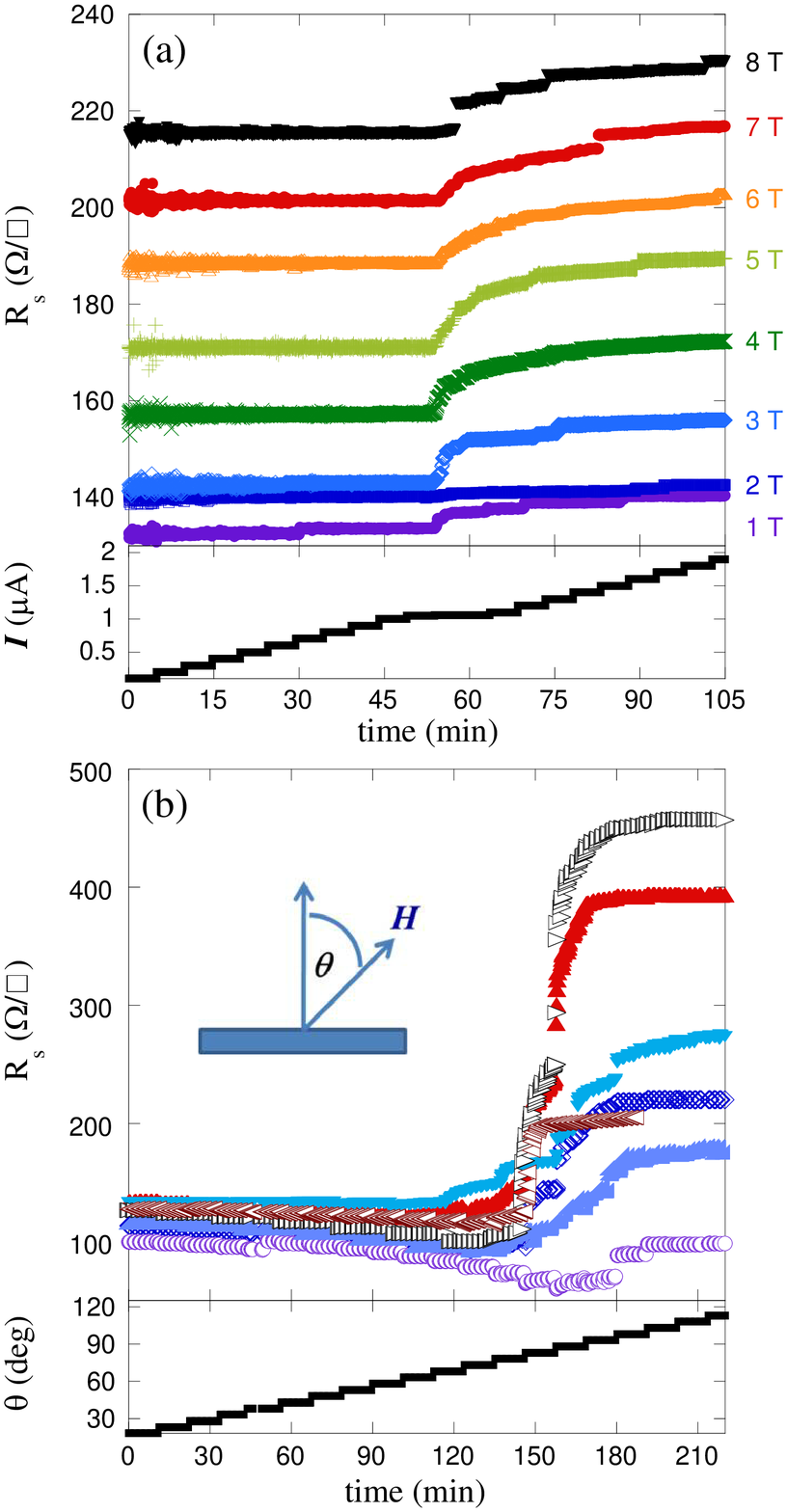}
\caption{(a) $R_s$ as a function of time at 5 K in a similar pattern as an in-plane magnetic field is applied in the sample plane and is increased in steps, and for each field, the current is increased in steps (as in Figure \ref{trapping}b). (b) $R_s$ as a function of time for several similar patterns as the angle $\theta$ between a magnetic field of 8 T and the normal to the interface is increased in steps.}
\label{trapping2}
\end{figure}

Figure \ref{trapping} presents the phenomenon of $R_s$ increase in connection with driving electrical currents and applying in-plane magnetic fields. Figure \ref{trapping}a shows the time dependence of $R_s$ at 5 K in a pattern with $d=5 \ \mu$m as an in-plane magnetic field is increased in steps. We see some increase of $R_s$ at low fields which cannot be distinguished from the increase observed at zero field; however, we note a significant enhancement of the effect for fields above 2 T. Figure \ref{trapping}b shows the time dependence of $R_s$ at 5 K in a similar pattern as the current is increased in steps (with smaller steps near 1 $\mu$A as shown in the inset). Here again, for low currents there is some increase of $R_s$ which cannot be distinguished from the observed increase in the zero-current limit; however, we note an accelerated increase in $R_s$ for currents above a certain current threshold of $\sim 1 \ \mu$A as shown in the inset for more patterns (black squares) and different cooling cycles of the same pattern (blue triangles). Heating is not involved in the observed $R_s$ increase as the same resistance is measured with low currents after the observed increase in $R_s$, as shown in Figure \ref{trapping}b. The systematic study of current induced $R_s$ increase was performed for patterns with $d=5 \ \mu$m. In wider patterns, the effect is significantly smaller. On the other hand, in patterns with $d=2 \ \mu$m, the $R_s$ increase may exceed 10 M$\Omega$ due to spontaneous increase alone which makes it more difficult to study systematically current- and field-induced effects.

Figure \ref{trapping2}a shows a combined effect of magnetic field and current at 5 K in a similar pattern. An in-plane magnetic field is applied in the sample plane and is increased in steps. For each field step, the time dependence of $R_s$ is measured as the current is increased in steps (with smaller steps near 1 $\mu$A as in Figure \ref{trapping}b). For easier comparison, we show the time dependence of $R_s$ for all the fields relative to the beginning of the current sweep. We note that the threshold current is insensitive to the field. Figure \ref{trapping2}b shows the effect of the angle $\theta$ between the field and the normal to the interface at 5 K for several similar patterns ($d=5 \ \mu$m). A field of 8 T is applied perpendicular to the sample plane and the time dependence of $R_s$ is measured as $\theta$ is increased in steps. The current used for this measurement was 1 $\mu$A. We note that the magnitude of the increase in $R_s$ is enhanced significantly as the field approaches the in-plane direction. We also note that the $R_s$ increase in the different patterns has a different magnitude despite the fact that their initial $R_s$ values were very similar (as well as their sheet carrier density ($n_s$) and the mobility ($\mu$)).

\begin{figure}[ht]
\includegraphics[scale=0.49, trim=100 0 100 0]{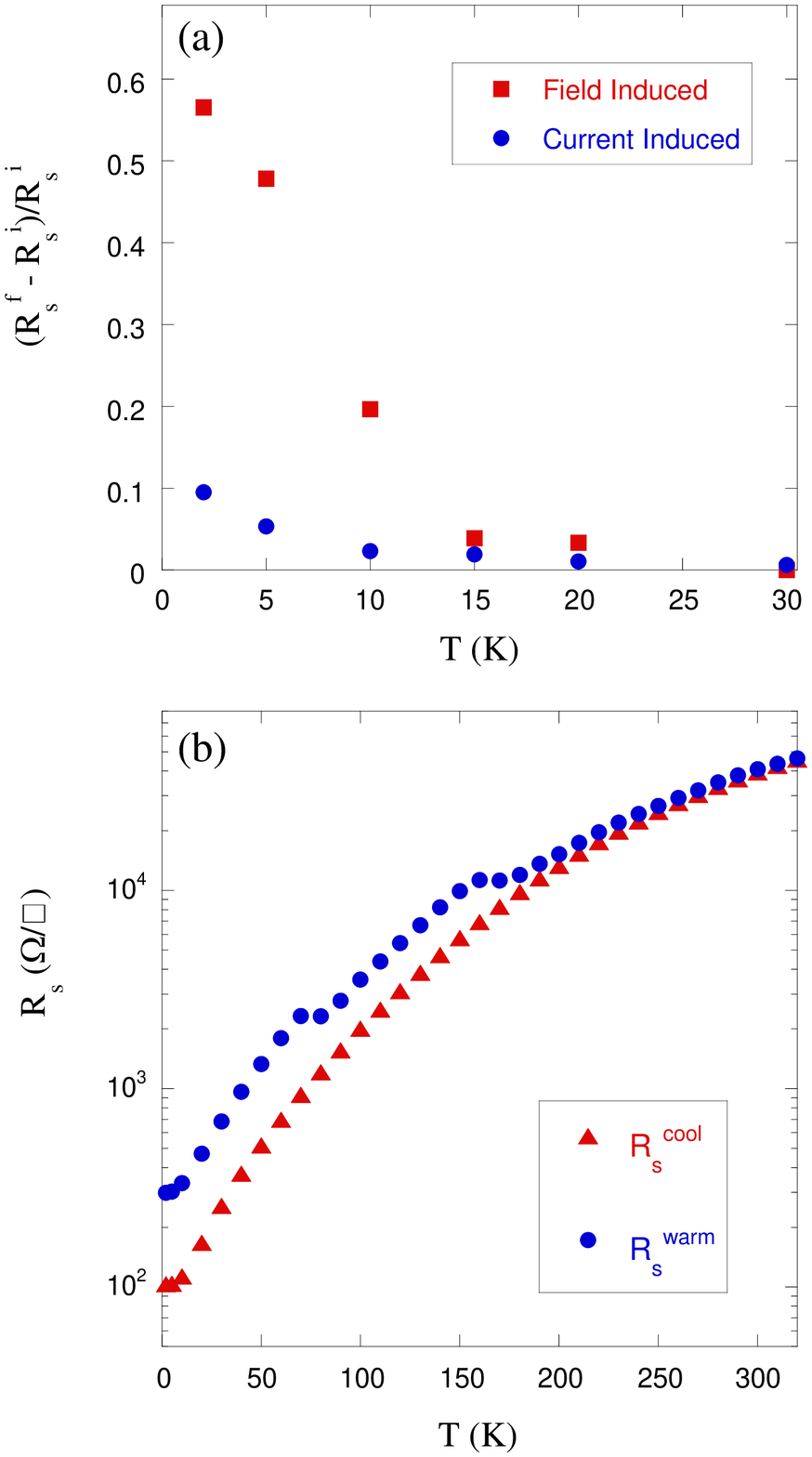}
\caption{(a) Temperature dependence of the relative increase in $R_s$ induced by in-plane fields or currents. $R^{i}_{s}$ is the initial $R_s$ and $R^{f}_{s}$ is $R_s$ after the induced increase. (b) Temperature dependence of $R_s$ during cooling (triangles) and during warming after an $R_s$ increase at low temperatures (circles).}
\label{Temperature}
\end{figure}

The observed current- and field-induced increases in $R_s$ are low-temperature phenomena. Figure \ref{Temperature}a shows the temperature dependence of the $R_s$ increase and we see that it becomes negligible above 30 K. The current- and field-induced increases in $R_s$ were achieved using the protocols shown in Figs. \ref{trapping}a and \ref{trapping}b.

The current- and field-induced increases in $R_s$ are reversible by warming. Figure \ref{Temperature}b shows $R_s$ during cooling from room temperature down to 2 K (triangles). At 2 K, $R_s$ is increased and then it is measured during warming (circles) in temperatures steps of ${\rm 10 \ K}$. The time interval between temperature steps is about 6 min and at each temperature, measurements are taken for 1 min.
The figure indicates accelerated conductivity recovery near 70 K and 160 K - these features are insensitive to the method used to increase $R_s$  at low temperatures. We note that puzzling transport properties were reported previously around these two temperatures \cite{experimental investigation}.

Based on measurements of tens patterns, we find that the magnitude of the $R_s$ increase depends strongly on the width of the current path and is very nonuniform. Patterns with wide current path (50 and 100 $\mu$m) do not exhibit field- and current-induced $R_s$ increases larger than a few percent. On the other hand, for a current path width of 5 $\mu$m, we observe occasionally $R_s$ increases larger than 100$\%$ which varies considerably between different patterns on the same sample and even between adjacent segments of the same pattern.

\begin{figure}[ht]
\includegraphics[scale=0.5, trim=100 0 100 0]{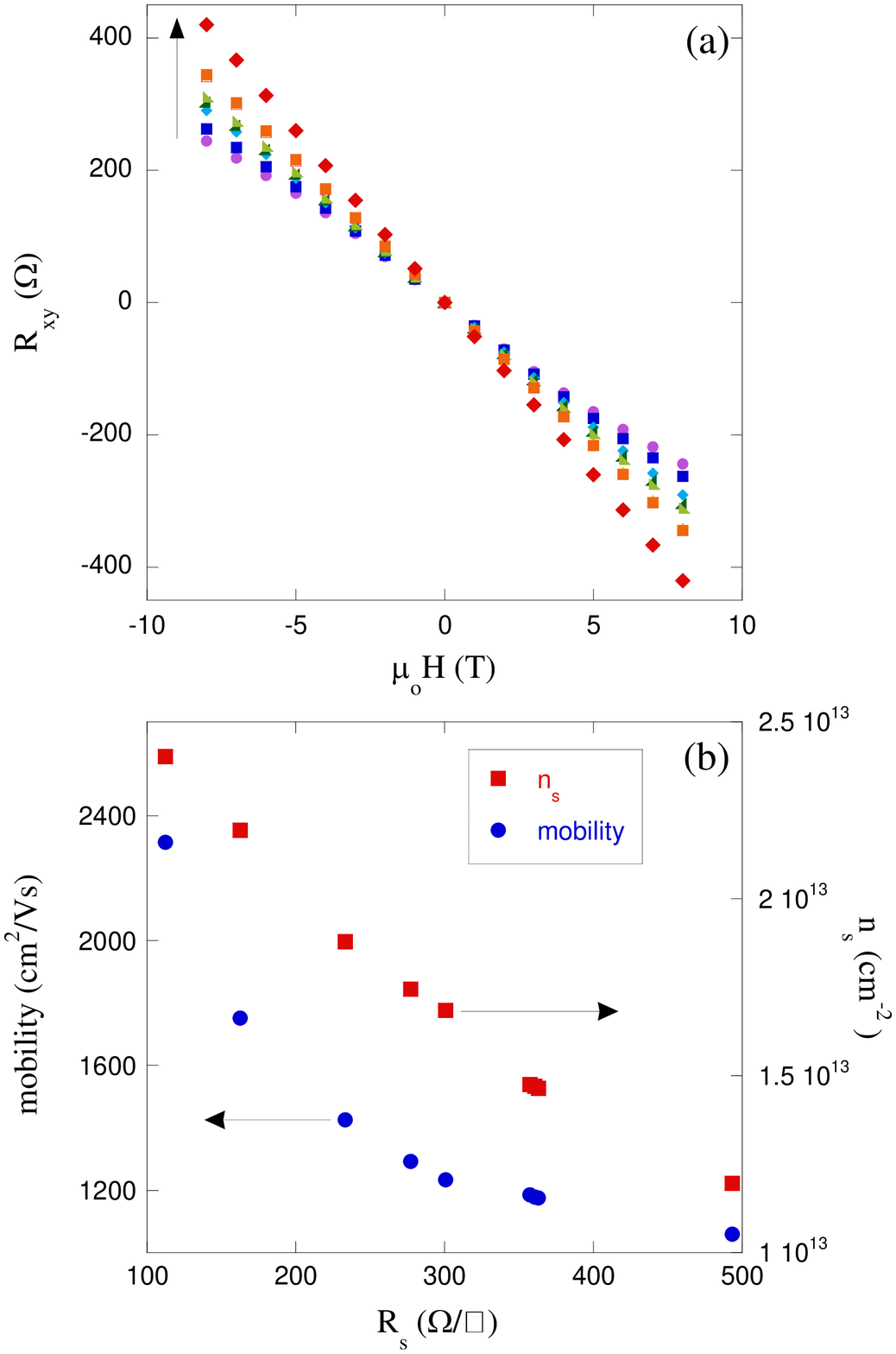}
\caption{(a) HE resistance vs perpendicular magnetic field for different
resistance steps at 2 K (see text). The arrow indicates the $R_s$-increase
direction. (b) Sheet carrier density and mobility as a function of
$R_s$ for the different resistance steps at 2 K.} \label{HE}
\end{figure}

To explore the changes in the transport properties
that occur concomitantly with the increase of $R_s$, we increase
$R_s$ of a given pattern ($d=5 \ \mu$m) in steps by applying an in-plane magnetic
field and/or a high electrical current for a time interval. For each
$R_s$, we measure the HE and the MR with a perpendicular field, and
the MR for a given field as a function of field orientation. The
latter is performed fast relative to the time scale of the field-induced effect to reduce the probability of $R_s$ increases during measurement. Figures \ref{HE}a and \ref{HE}b
show the HE resistance ($R_{xy}$) for the different values of $R_s$
and the dependence of $n_s$ and $\mu$ on $R_s$, at 2 K. We note that as $R_s$ increases, $n_s$
and $\mu$ decrease. Similar correlations between $R_s$, $n_s$ and
$\mu$ are observed by us in other samples and are also reported for
gated samples \cite{dominant mobility}.

The fact that at low temperatures $R_s$ always increases and that its increase is accompanied by a decrease in $n_s$ may suggest charge trapping as a possible mechanism. One may think that current and field may enhance trapping by applying a Lorentz force on the charge carriers. However, $R_s$-increase experiments where the directions of the current and the field were kept constant showed no dependence on the current and field polarity nor on the angle between the current and the in-plane field. Although we cannot determine based on our measurements where the charges might be trapped, the fact that the $R_s$ increase can be much larger when the current path is few microns wide suggests that the trapping is correlated on a similar length scale. The accelerated recovery of $R_s$ upon warming near ${\rm 70 \ K}$ and ${\rm 160 \ K}$ suggests two distinct trapping potentials. The detailed study of the relaxation of $R_s$ in the vicinity of these two temperatures will be reported elsewhere.

\begin{figure}[ht]
\includegraphics[scale=0.6, trim=100 0 100 0]{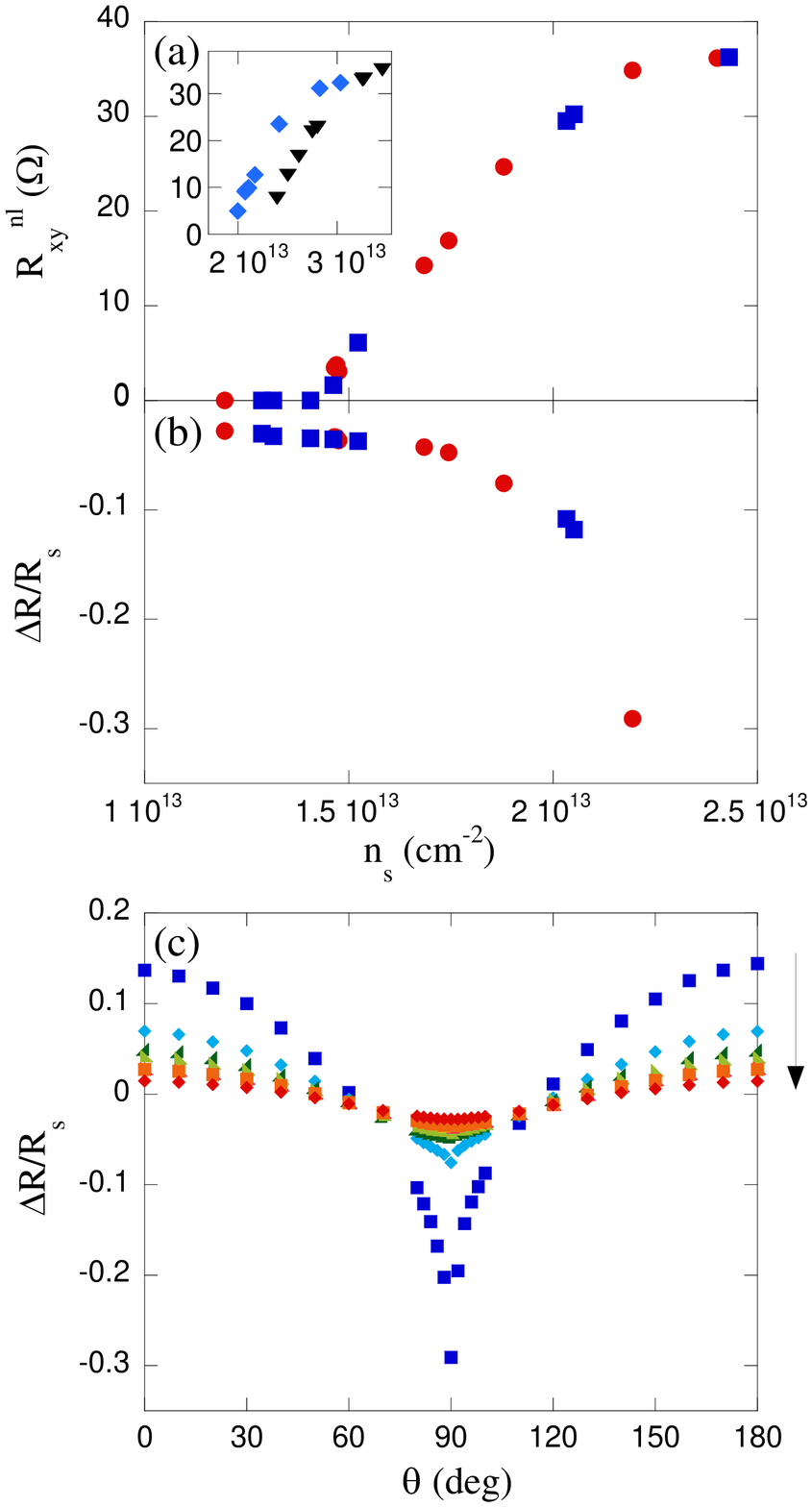}
\caption{(a) The non-linear part of the HE resistance ($R^{nl}_{xy}$) as a function
of $n_s$. Solid squares and solid circles represent different sets of
$R_s$ increases using the same pattern ($d=5 \ \mu$m). Inset: $R^{nl}_{xy}$ vs $n_s$ for two other patterns. (b) The magnetoresistance $\Delta$R/$R_s$=$\frac{R_s(H)-R_s(H=0)}{R_s(H=0)}$ at
$\theta$=$90^\circ$ as a function of $n_s$. Symbols are as in panel (a).
(c) The magnetoresistance at 2 K as a function of $\theta$ with a field of
8 T for the resistance steps as shown in Figure \ref{HE}a. The arrow
indicates the $R_s$-increase direction.} \label{steps}
\end{figure}

\begin{figure}[ht]
\includegraphics[scale=0.47, trim=100 0 100 0]{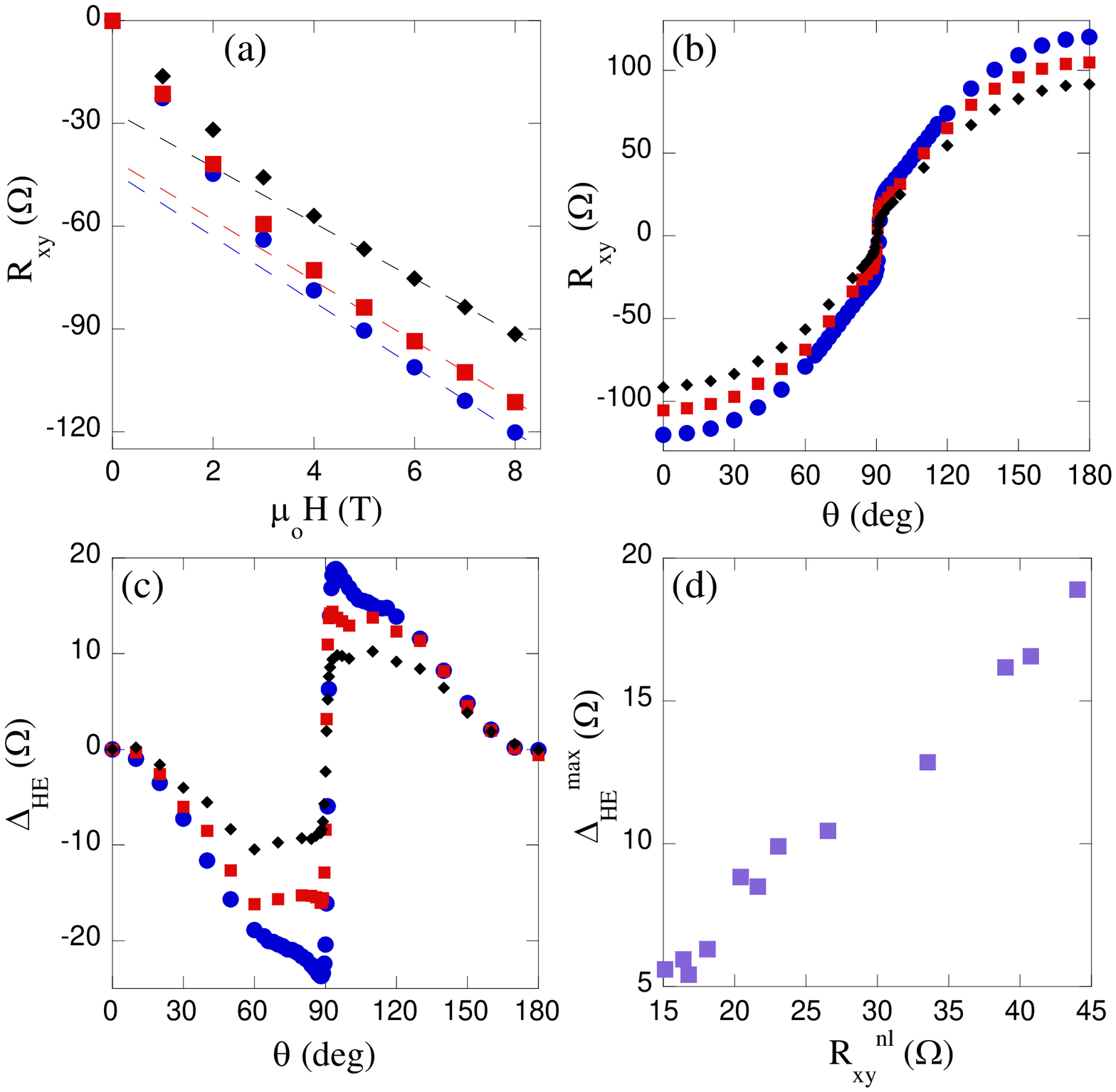}
\caption{(a) The HE resistance vs perpendicular magnetic field for
three different patterns with $d=50 \ \mu$m marked with
different symbols at 2 K. The dashed lines are extrapolations of high-field
HE. (b) The HE resistance as a function of $\theta$ with a field of
8 T for the three different patterns at 2 K. (c) The deviation of the HE
resistance angular dependence from a $\cos \theta$ behavior,
$\Delta_{HE}$=$R_{xy}(\theta)-R_{xy}(\theta=0)\cos\theta$, as a
function of the angle $\theta$. (d) The maximal value of
$\Delta_{HE}$, $\Delta^{max}_{HE}$, as a function of the non-linear
part of the HE resistance, $R^{nl}_{xy}$, in nine different patterns ($d=50 \ \mu$m).} \label{EHE}
\end{figure}

We now turn to explore the changes in the magnetotransport
properties that occur as $R_s$ increases. Figure \ref{HE}a shows
that, while the HE resistance of the as-cooled sample is nonlinear
in field, the nonlinearity gradually vanishes with increasing
$R_s$. We extrapolate the high-field linear dependence of the HE
resistance to low fields and define its zero field value as the
nonlinear component of the HE resistance denoted as $R^{nl}_{xy}$.
Figure \ref{steps}a shows $R^{nl}_{xy}$ as a function of $n_s$ and
we observe its gradual decrease with decreasing $n_s$, starting at $2.4 \times 10^{13} \ {\rm cm^{-2}}$ until it vanishes for $n_s\sim 1.4 \times 10^{13} \ {\rm cm^{-2}}$. We note, however, that this value is sample dependent and seems to be correlated with the as-cooled value of $n_s$. As shown in the inset, for a pattern with as-cooled $n_s$ of $3 \times 10^{13} \ {\rm cm^{-2}}$, the nonlinear part vanishes for $n_s\sim 1.9 \times 10^{13} \ {\rm cm^{-2}}$, and for a pattern with as-cooled $n_s$ of $3.5 \times 10^{13} \ {\rm cm^{-2}}$, the nonlinear part vanishes for $n_s\sim 2.3 \times 10^{13} \ {\rm cm^{-2}}$.

Concomitantly with the vanishing of $R^{nl}_{xy}$, the sharp
negative dip in the MR angular dependence obtained near
$\theta$=$90^\circ$ and linked to magnetism
\cite{antiferromagnetism,antisymmetry,2D_3D} is rapidly suppressed.
Figure \ref{steps}c shows the MR vs $\theta$ for different
resistance steps at 2 K. The rapid suppression of the negative MR at
$\theta$=$90^\circ$ as $n_s$ decreases is shown in Figure
\ref{steps}b.

The suppression of $R^{nl}_{xy}$ with increasing $R_s$ would suggest
suppression of magnetism if the nonlinearity can be attributed to
anomalous HE (AHE). However, nonlinear HE at $\theta = 0^\circ$ can
be attributed to the contribution of two (or more) bands \cite{dominant
mobility,SDH Oscillations}. Measuring the HE as a function of
$\theta$ may distinguish between the two scenarios. While the
ordinary HE (OHE) in the two-band scenario follows a $\cos \theta$
angular dependence (even when it exhibits a nonlinear field
dependence at $\theta$=$0^\circ$), the contribution of the AHE may
show deviations from a $\cos \theta$ angular dependence when the
magnetization is not parallel to the applied magnetic field due to
magnetic anisotropy or when the magnitude of the magnetization
varies with $\theta$.

Figure \ref{EHE}a shows the HE of three different patterns that exhibit nonlinear field dependence at $\theta$=$0^\circ$ at 2 K. For the same patterns, we show in Figure \ref{EHE}b the angular dependence of the HE at 2 K, which indicates clear deviations from a $\cos \theta$ behavior. To
highlight the deviations, we present in Figure \ref{EHE}c the
difference between $R_{xy}(\theta)$ and the expected angular dependence of
an OHE with the same value for $\theta=0^\circ$ ($\Delta_{HE}$). We
see that the deviations are particularly large near $\theta$=$90^\circ$. The data shown in Figure \ref{EHE}
are for patterns with $d=50 \ \mu$m which exhibit relatively small $R_s$ increases during the angular-dependent measurements (less than 3$\%$). However, the features shown in the figure are observed in tens patterns of different samples with various current path widths, including in patterns with $d=5 \ \mu$m which exhibit large $R_s$ increases; although for these patterns, the data are significantly affected by the field-induced increase in $R_s$ as $\theta$ approaches $90^\circ$.

Figure \ref{EHE}d shows for nine different patterns with similar as-cooled $n_s$ the maximum value
of $\Delta_{HE}$ ($\Delta^{max}_{HE}$) as a function of
$R^{nl}_{xy}$. The figure indicates that there is a clear
correlation between the deviations from a $\cos \theta$ behavior and
the nonlinearity of the HE. This correlation is expected if the
source of the nonlinearity is magnetism. It is less expected if the
source of the nonlinearity is a two-band contribution where no deviation
from a $\cos \theta$ behavior is expected.



The interplay between the nonuniform $R_s$ increase and the
dramatic changes in magnetotransport properties suggests that charge
trapping, which decreases carrier density and increases disorder,
suppresses magnetism in this system. The nonuniform nature of the
magnetism we find is consistent with recent experiments which
present unequivocal evidence for magnetism
\cite{Coexistence2,Coexistence3}. On the other hand, we note that
whereas the angular dependence of the MR and the HE suggests
perpendicular magnetic anisotropy \cite{antisymmetry}, these reports
find mainly in-plane magnetic moment. The suppression of
magnetization may be linked to the decrease in $n_s$ or to the
increase in disorder. The fact that the value of $n_s$ at which
$R^{nl}_{xy}$ disappears varies between samples may imply that the
existence of magnetic properties is not determined only by $n_s$.

In conclusion, we identify low-temperature $R_s$ increases in connection with driving an electrical current higher than a certain threshold through the pattern and/or applying a strong
enough in-plane magnetic field. Irrespective of the method used to
increase $R_s$, as the sample is warmed up, $R_s$ recovers to its
original value with accelerated recovery near 70 K and 160 K. The
$R_s$ increase is accompanied by a decrease in carrier density and
mobility, and since it is nonuniform, it increases the
nonhomogeneity of the sample. In addition, we find a clear link between the
$R_s$ increase and the suppression of magnetotransport properties
which can be linked to magnetism. The existence of the $R_s$-increase mechanism and its effect on magnetotransport is important both for future design of submicron devices and for elucidating the transport and magnetic properties of the LAO/STO system.

\section{acknowledgements}
L.K. acknowledges support by the German Israeli Foundation (Grant
No. 979/2007) and by the Israel Science Foundation founded by the
Israel Academy of Sciences and Humanities (Grant No. 577/07). We
acknowledge the contribution of J. Mannhart, R. Jany and S. Paetel
who provided the samples used for this research. We acknowledge
useful discussions with E. Shimshoni.

\end{document}